\documentclass[journal]{IEEEtran}
\usepackage{cite}
\usepackage{amsmath,amssymb,amsfonts}
\usepackage{graphicx}
\usepackage{textcomp}
\usepackage{multicol}
\usepackage{balance}
\usepackage{soul}
\usepackage{multirow}
\usepackage{makecell}
\usepackage[table,xcdraw]{xcolor}
\usepackage{tikz}
\usetikzlibrary{shapes}
\usepackage{subfigure}
\usepackage[ruled,vlined,linesnumbered]{algorithm2e}
\usepackage[hyphens]{url}
\def\BibTeX{{\rm B\kern-.05em{\sc i\kern-.025em b}\kern-.08em
    T\kern-.1667em\lower.7ex\hbox{E}\kern-.125emX}}

\begin{document}

\title{A Survey on Federated Learning for the Healthcare Metaverse: Concepts, Applications, Challenges, and Future Directions}
\author{}
\author{Ali Kashif Bashir,~\IEEEmembership{Senior Member,~IEEE}, Nancy Victor,~\IEEEmembership{Member,~IEEE}, Sweta Bhattacharya, Thien Huynh-The, Rajeswari Chengoden, Gokul Yenduri,~\IEEEmembership{Student Member,~IEEE}, Praveen Kumar Reddy Maddikunta, Quoc-Viet~Pham, Thippa Reddy Gadekallu,~\IEEEmembership{Senior Member,~IEEE}, and Madhusanka~Liyanage,~\IEEEmembership{Senior Member,~IEEE} 
\thanks{Corresponding Author: Thippa Reddy Gadekallu}
\thanks{Ali Kashif Bashir is with Manchester Metropolitan University, United Kingdom (email: Dr.alikashif.b@ieee.org)}
\thanks{Nancy Victor, Sweta Bhattacharya, Rajeswari Chengoden,  Gokul Yenduri, Praveen Kumar Reddy Maddikunta are with the School of Information Technology, Vellore Institute of Technology, Tamil Nadu-632014, India (Email: \{nancyvictor, sweta.b, rajeswari.c, gokul.yenduri, praveenkumarreddy\}@vit.ac.in). }
\thanks{Quoc-Viet~Pham is with School of Computer Science and Statistics, Trinity College Dublin, The University of Dublin, Ireland (email: viet.pham@tcd.ie).}
\thanks{Thippa Reddy Gadekallu is with the School of Information Technology and Engineering, Vellore Institute of Technology, Vellore 632014, India as well as with DeparFtment of Electrical and Computer Engineering, Lebanese American University, Byblos, Lebanon (Email: thippareddy.g@vit.ac.in)}
\thanks{Thien Huynh-The is with the Department of Computer and Communication Engineering, Ho Chi Minh City University of Technology and Education, Vietnam (email: thienht@hcmute.edu.vn).}
\thanks{Madhusanka~Liyanage is with School of Computer Science, University College Dublin, Ireland (email: madhusanka@ucd.ie).}
}

\IEEEtitleabstractindextext{%
\begin{abstract}
Recent technological advancements have considerately improved healthcare systems to provide various intelligent healthcare services and improve the quality of life. Federated learning (FL), a new branch of artificial intelligence (AI), opens opportunities to deal with privacy issues in healthcare systems and exploit data and computing resources available at distributed devices. Additionally, the Metaverse, through integrating emerging technologies, such as AI, cloud edge computing, Internet of Things (IoT), blockchain, and semantic communications, has transformed many vertical domains in general and the healthcare sector in particular. Obviously, FL shows many benefits and provides new opportunities for conventional and Metaverse healthcare, motivating us to provide a survey on the usage of FL for Metaverse healthcare systems. First, we present preliminaries to IoT-based healthcare systems, FL in conventional healthcare, and Metaverse healthcare. The benefits of FL in Metaverse healthcare are then discussed, from improved privacy and scalability, better interoperability, better data management, and extra security to automation and low-latency healthcare services. Subsequently, we discuss several applications pertaining to FL-enabled Metaverse healthcare, including medical diagnosis, patient monitoring, medical education, infectious disease, and drug discovery. Finally, we highlight significant challenges and potential solutions toward the realization of FL in Metaverse healthcare. 
\end{abstract}


}

\maketitle

\IEEEdisplaynontitleabstractindextext

\IEEEpeerreviewmaketitle

\section{Introduction} 
Healthcare is a system in which people are paid attention to improving their physical, mental, and cognitive health by means of diagnosis and prognosis measures. It is delivered by healthcare professionals, lab technicians, and medical staff. The conventional healthcare system served the community abiding by medical policies such as Health Insurance Portability and Accountability Act (HIPAA), World Health Organization (WHO), etc. One of the latest trends in healthcare is the use of wearable devices to collect health data. It enables continuous data tracking and immediate healthcare interventions. Internet of Medical Things (IoMT) is a sub-area of IoT that deals with the collection of medical devices and applications that connect through computer networks. 
The data collected with the aid of wearable devices and IoMT are used for monitoring and further treatments. Privacy is one of the major concerns in wearable technologies \cite{dwork2006differential}. Wearable technologies can give insights about the individual’s health as the data are shared over the internet.
 Metaverse is a confluence of multiple technologies like Artificial Intelligence (AI), Virtual Reality (VR), Augmented Reality (AR), Internet of Medical Things (IoMT) etc. The health and fitness of the patients can be monitored and further improved through the use of Metaverse technologies \cite{wang2022development,zhou2023resource, zhou2022joint, zhou2022mobile, zhou2022resource}.  Metaverse in healthcare helps in surgical simulation, diagnostic imaging, rehabilitation, and patient health management\cite{bansal2022healthcare}. Moreover, Metaverse has the ability to motivate the patients to persistently do the required physical movements according to the treatment procedures. VR-based treatments offer a drug-free method for trauma-related anxiety.

The existing healthcare system that utilizes the Metaverse stores the data in a private or central cloud. Intelligent decisions are derived from the data which is being shared with the individual nodes \cite{zhang2023multi}. 
Centralized decision-making is one of the issues identified in the Metaverse for healthcare. The data are collected from multiple hospitals and stored in a central repository. The Metaverse users majorly use body-worn devices, and much of the data collected is pertaining to the individual’s behavior and the surroundings.  Data collected from wearable devices are sent to the computer and to the cloud. In all such cases, the data is shared in a public cloud and hence there is a huge chance for the privacy to be compromised. Latency is the time taken for the data to be transferred from one point to the other \cite{gupta2023understanding}. A number of the devices are connected and the users are connected remotely in the Metaverse environment. Providing fast, jitter-free, the low latent end-to-end connection is a significant challenge to be tackled in the Metaverse applications.

The healthcare Metaverse application gets benefits from Federated Learning (FL) which enables individual hospitals to collaborate and learn from the shared predictive models. FL allows on-device learning which helps to overcome the issues like data privacy, data security, and high data latency. The distributed processing nature of FL enables the healthcare Metaverse gradually learn from the heterogeneous data \cite{han2023application}.

In \cite{qiao2022sprechd} medical Metaverse is used to diagnose fetal Congenital Heart Disease (CHD). Fetal ultrasound is obtained through ECG screening of pregnant women. The ultrasound dataset can be stored and thereafter AI models are deployed constituting of XR/VR to construct holographic fetal heart. The cardiologists from different regions view the holographic fetal heart, refer to the predictive results of the AI model and conduct real-time consultation. The authors in \cite{song2022Metaverse} provide perspectives on the digital twin-based personal healthcare framework that is used for the assessment and maintenance of personal health. 
In \cite{wu2020effect}, the authors have developed a VR based system and implemented the same for providing training to nurses and medical interns. It is important to mention here that collaborations among pharmaceutical institutions often make companies get susceptible to data privacy compromises \cite{chen2021fl}. Horizontal FL provides an efficient solution to promote collaboration ensuring data privacy is retained. In \cite{dou2021federated}, the authors have proposed a framework to maximize the potential of AI modes by scaling the medical training dataset and enriching the heterogeneity of patient data distribution across the world. The proposed framework also shows interest in preserving the privacy of the data by fusing FL to the existing methodologies. 
To the best of our knowledge, there has not been a study that discusses the contribution of FL in healthcare Metaverse. Thus our study is an informative survey that presents potential applications of integrating FL which would enhance privacy and in association to this, Metaverse would enhance the effectiveness in providing healthcare services. Table 1 presents the contribution of this paper. 

The significant contributions of this paper are highlighted below:
\begin{enumerate}
    \item To understand the differences between conventional healthcare and the healthcare Metaverse
    \item To identify the applications of FL-enabled healthcare Metaverse
    \item To highlight the potential challenges of implementing FL in the healthcare Metaverse and present possible future directions
\end{enumerate}

\begin{table*}[h!]
\centering
\caption{A comparison of this survey with existing surveys}
\label{tab:comparision}
\begin{tabular}{|l|l|l|p{.3cm}p{2cm}p{2cm}p{2cm}p{2cm}|}
\hline
\multicolumn{1}{|c|}{\multirow{2}{*}{\textbf{Ref.No}}} &
  \multicolumn{1}{c|}{\textbf{FL}} &
  \multicolumn{1}{c|}{\textbf{Metaverse}} &
  \multicolumn{5}{c|}{\textbf{Applications}} \\ \cline{2-8} 
\multicolumn{1}{|c|}{} &
  \multicolumn{1}{c|}{\textbf{}} &
  \multicolumn{1}{c|}{\textbf{}} &
  \multicolumn{1}{c|}{\textbf{Medical Diagnosis}} &
  \multicolumn{1}{c|}{\textbf{Patient Monitoring}} &
  \multicolumn{1}{c|}{\textbf{Research \& Education}} &
  \multicolumn{1}{c|}{\textbf{Infectious Diseases}} &
  \multicolumn{1}{c|}{\textbf{Drug Discovery}} \\ \hline
\cite{ qiao2022sprechd} &
   &
  {$\checkmark$} &
  \multicolumn{1}{l|}{$\checkmark$} &
  \multicolumn{1}{l|}{} &
  \multicolumn{1}{l|}{} &
  \multicolumn{1}{l|}{} &
   \\ \hline
\cite{song2022Metaverse} &
   &
  {$\checkmark$} &
  \multicolumn{1}{l|}{} &
  \multicolumn{1}{l|}{$\checkmark$} &
  \multicolumn{1}{l|}{} &
  \multicolumn{1}{l|}{} &
   \\ \hline
\cite{wu2020effect} &
   &
  {$\checkmark$} &
  \multicolumn{1}{l|}{} &
  \multicolumn{1}{l|}{} &
  \multicolumn{1}{l|}{$\checkmark$} &
  \multicolumn{1}{l|}{} &
   \\ \hline
\cite{chen2021fl} &
  {$\checkmark$} &
   &
  \multicolumn{1}{l|}{} &
  \multicolumn{1}{l|}{} &
  \multicolumn{1}{l|}{} &
  \multicolumn{1}{l|}{} &{$\checkmark$} \\ \hline
\cite{ dou2021federated} &
  {$\checkmark$} &
   &
  \multicolumn{1}{l|}{} &
  \multicolumn{1}{l|}{} &
  \multicolumn{1}{l|}{} &
  \multicolumn{1}{l|}  {$\checkmark$} &
   \\ \hline
Our Paper &
  {$\checkmark$} &
  {$\checkmark$} &
  \multicolumn{1}{l|}{$\checkmark$} &
  \multicolumn{1}{l|}{$\checkmark$} &
  \multicolumn{1}{l|}{$\checkmark$} &
  \multicolumn{1}{l|}{$\checkmark$} &
  {$\checkmark$} \\ \hline
\end{tabular}%
\end{table*}

The rest of the paper is structured as follows. Section II highlights the aspects of conventional healthcare, Metaverse healthcare and the significance of FL in conventional healthcare, whereas section III provides a detailed description about the benefits of FL in the healthcare Metaverse. Section IV discusses about the various applications of adapting FL for the healthcare Metaverse. The significant challenges and possible future directions are presented in section V and section VI concludes the paper. 
\section{Background}
This section presents the significant aspects of conventional healthcare and the healthcare Metaverse. The key benefits of the healthcare Metaverse are also discussed in this section. A detailed description about the significance of implementing FL in the healthcare Metaverse is provided along with a discussion on the healthcare Metaverse platforms. 
\subsection{Conventional Healthcare and Healthcare Metaverse}
In the past, before the development and popularity of Internet-of-Things (IoT) systems for medical and healthcare, patients have to spend more time and cost for clinics-related services \cite{baucas2023federated}. However, with several advantages for patients, hospitals, and medical centers, such as automatic health monitoring and remotely collaborative medical diagnosis, many healthcare and medical systems based on IoT with cutting-edge technologies (such as big data, artificial intelligence, and fifth generation) have completely revolutionize the healthcare industry to create more beautiful things all over the globe.  In many IoT-based healthcare systems, healthcare and medical data (e.g., sequential data as electroencephalogram, high-dimensional data as ultrasound images, and documents as electronics health records) are collected using various smart sensor-based devices (e.g., biosensors, wearable devices, smart gadgets, and medical equipment) with internet connection \cite{singh2023privacy}. The data is then stored in the local and its copy should be transmitted to the data center for cloud storage in the case of direct acquisition from devices. At the data center, the data is processed and analyzed to extract meaningful and relevant information automatically with diversified artificial intelligence (AI) tools, including feature engineering and machine learning (ML) algorithms. Finally, the results obtained by AI in service-oriented computation tasks are communicated to either the individual patients and the health service providers, wherein some primitive results can help doctors and health experts to make the final decision of medical diagnosis and plan the most appropriate treatment. It should be noted that all the stages, starting from sensing to data acquisition and processing, require security mechanisms, from the physical layer to service and application layers, to protect sensitive health data, high-level information, and diagnosis result from cyberattacks \cite{wen2023health}. 

There are four layers in general IoT-based healthcare and medical systems: sensing layer, network layer, processing layer, and application layer. 
\begin{itemize}
\item \textit{Sensing layer}: As the patient-nearest module, this layer is responsible for data acquisition, which is arranged by several wearable sensors to measure physiological parameters from human body (e.g., temperature, blood pressure, and glucose level). The number and types of sensors in use may vary upon the applications for consideration and the quality of providing services. Sensing usually plays an important role in various healthcare applications because the sensed value acts as the input for processing and may affect to the final diagnosis decision hereafter. Depending on the purpose, there are two common options of sensory data storage: cloud-based and server-based infrastructures \cite{tian2023biocompatible}.
\item \textit{Networking layer}: This layer is responsible for communication and security related activities, that means, selecting appropriate data transmission protocols to transfer data from source (e.g., heterogeneous sensors, devices, and medical equipment) to destination (e.g., server and cloud) and encryption mechanisms to protect the data privacy. There are several communication protocols, but they are mostly categorized into three fashions depending on the requirements of reliability and effectiveness: centralized, decentralized, and hierarchical. During the data transmission phase, some advanced security schemes can be developed to detect and recognize cyberattacks, such as ransomware, data breaches, DDoS attacks, insider threats, and fraud scams \cite{gupta2023fedcare}.
\item \textit{Processing layer}: Data management and computation related activities are considered in this layer, in which various data analytic techniques are performed to extract relevant and meaningful information from raw sensory data. For data management, three major aspects are investigated, including storage type (centralized and distributed schemes), processing medium (in-network-based, base-station-based, and cloud-based data processing approaches), and data integration (individual and combined mechanisms). For computation, depending on the data types, applications, and requirements, selecting an appropriate modeling method is crucial. For instance, the static modeling methods can generate an application-specific model to deal with the same disease of different patients. Compared with static modeling, the dynamic modeling can achieve real-time prediction and monitoring in various complicated health scenarios, in which the AI technologies with ML algorithms can be applied to enhance the overall performance of systems \cite{qadri2020future}. 
\item \textit{Application layer}: This layer maganes all control/interaction/actuation and application related activities, wherein the patients can receive feedback and the clinicians can view primitive analysis and diagnosis results from the processing layer \cite{sultana2019choice}. In many automated healthcare systems, all potential solutions and recommendations are given by systems based on the learned models, whereas the manual mode provides suggestions and recommendations to doctors to make the decision based on the combination of his experiences and patients’ conditions. Besides, a verity of applications are developed with user interface as mobile and desktop apps, in which the built-in functionalities and services may be easy for patients and advanced for medical and wellness experts. It should be noted that the complexity of an IoT-based healthcare system mostly depends upon the number of services and the complication of functionalities.
\end{itemize}

While Metaverse is unfolding to many industrial domains, it presents many opportunities in the healthcare sector that combines the cutting-edge technologies like AI, VR, IoT, intelligent cloud, collaborative robotics, edge and quantum computing to provide new directions and breathtaking experiences to healthcare. To exactly specify the definition of Metaverse Healthcare, it should interact with VR/AR glasses facilitated by Internet-of-Medical/Health-Things \cite{zikas2023mages}. AR manipulates visual elements and graphical characters to transmute the real world by enabling users to view their three-dimensional (3D) surroundings through AR devices, such as smart glasses. VR creates a fully computer-generated digital world that users can immerse with VR headsets, gloves, controller, and other digital sensors. Since healthcare and medical data are at high risk of being stolen due to its storage on centralized servers, service providers realized that they should build an effective and secure health diagnosis platform to deal with a daily increasing caseload. In this context, the Metaverse is already showing some significant benefits in digital therapeutics that VR and AR technologies provide medical applications and services with the support of haptic sensor and 3D interaction \cite{lim2022realizing}. Fig.~\ref{fig:FL Implications} depicts the upcoming implications of the Metaverse in Healthcare.  Some key benefits of Metaverse Healthcare are foreseeable to revolutionize the whole healthcare industry as follows:
\begin{figure*}[h!]
	\centering
	\includegraphics[width=\linewidth]{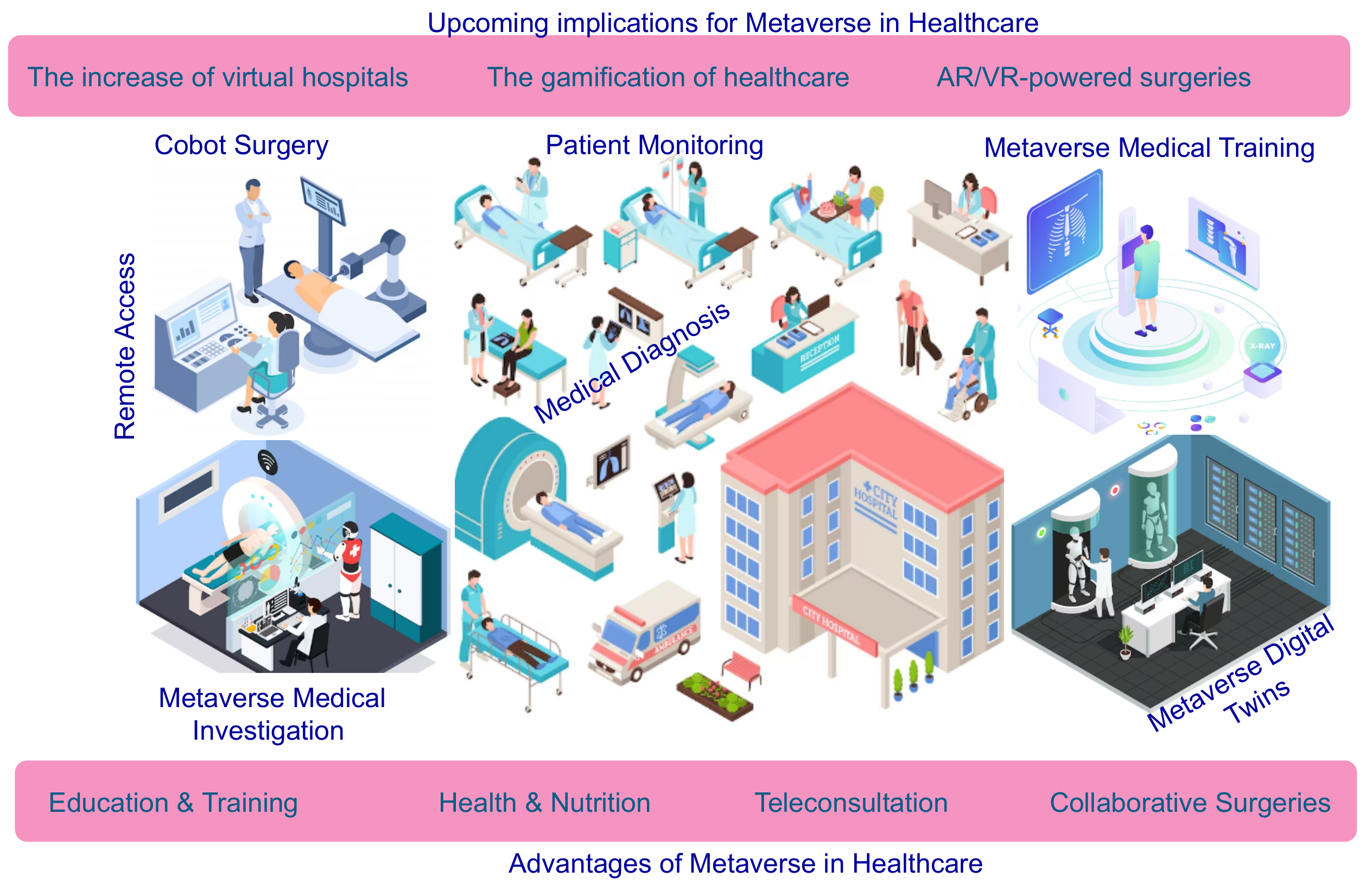}
	\caption{Upcoming implications of the Metaverse in Healthcare.}
	\label{fig:FL Implications}
\end{figure*}
\begin{itemize}
\item \textit{Medical training and education}: The Metaverse opens a new era for the healthcare industry, wherein cognitive therapy and rehabilitation are considered for development as built-in services in the Metaverse. Service providers efficiently interact with patients and other clinicians for remotely collaborative diagnosis, stimulate effects of a proposed treatment, and experience more personalized technologies for better diagnosis \cite{sandrone2022medical}. Within the Metaverse, medical students and trainees can be provided a deep understanding about human physiology over interactive holographic projections \cite{skalidis2022cardioverse}.
\item \textit{Change people lifestyle}: Digital platforms based on the Metaverse allows users to have an immersive experience by enticing a feeling of their all-time presence in connected virtual worlds. As one of some outstanding long-term effects of the Metaverse relating to mental health, psychiatrists and psychotherapists could exploit the immersive experiences in the Metaverse to treat several challenging issues, such as eating disorders, anxiety disorders, delusions, psychosis, and hallucinations \cite{koohsari2023metaverse}. Furthermore, gamification and personalization in the Metaverse can change the current lifestyle of people. For example, both the professionals and patients are recommended to follow a healthy lifestyle routine and get incentives in the form of tokens or rewards. 
\item \textit{Connected community}: Last but not least, the Metaverse is expected to be an ideal connected hub for people to socialize in a more and more digitized community \cite{musamih2022nfts}. In the Metaverse Healthcare, patients who have the same health issues or needed sesha, can share experiences, give advises, and support others for the better in the long run. Hospital staffs can resolve problems by keeping track of the time and workload seamlessly via a virtual space but using data collected from the real world. In other words, the Metaverse evolves the next generation of health facilities, wellness institutions, and medical centers which all are housed in a single Metaverse.
\end{itemize}

\subsection{FL in Conventional Healthcare}
As computer software and hardware technologies advance quickly, more and more healthcare data are becoming available from patients, healthcare organizations, the pharmaceutical industry, and insurance companies, among other sources. Data science technologies now have an improbable chance to get insights from data and raise the standard of healthcare services because of this access \cite{ali2022federated}. However, the acquisition of enormous, diversified, centrally stored healthcare datasets is almost impossible due to strict privacy laws and data ownership considerations. On the other hand, the AI model demands an increasing amount of healthcare data in order to provide better decisions. FL offers promising solutions to this issue, and the advantages of adapting FL over conventional methods are discussed below.
\begin{itemize}
\item \textit{Privacy}:
 FL offers training without compromising privacy of healthcare data. FL has enormous potential for connecting disparate healthcare data sources while protecting privacy. FL uses a central server to train a common global model while preserving all the sensitive data at the local institutions where it belongs \cite{patel2022adoption}. Thus, FL enables healthcare organisations to create reliable and trustworthy models by participating in collaborative training without disclosing data to third parties. As a result, it provides new opportunities for industry and research and enhances health care all around the world. FL can have positive impact on almost all stakeholders and the entire treatment cycle, better diagnostic tools for clinicians, including improved medical image analysis, collaborative and accelerated drug discovery, true precision medicine by assisting in the identification of similar patients, and  which reduces cost and time-to-market for pharmaceutical companies \cite{sadilek2021privacy}. 
 \item \textit{Reduced Computational Cost and Power Consumption}:
 The accuracy of AI models will increase as more data is gathered. However, if all the healthcare data is kept in one location, training a model will take more time and resources \cite{lu2022personalized}. With FL, there is no need to store data in a certain place. Instead, a global model can be trained based on the parameters shared by the local models. When compared to conventional models, FL does not require any personal data from the local models, which lowers the cost of computational resources and lowers power usage \cite{ng2021federated}.
 \item \textit{Personalization}: FL also allows local model to train at local network, and the cloud for global model aggregation. The distributed datasets, which may vary across various local nodes, are the major source of data used for training at the cloud server. The shared model may not work well for a specific user since it purely accounts for the traits that all health users have in common \cite{rehman2022federated}. As a consequence, each user incorporates the learned global model with his own unique health data to provide personalized health monitoring. The local devices get the global model from the cloud and train their unique model in order to capture the tailored services. Additionally, a balanced dataset is created for every device using the global model from the cloud, reducing the disparity between individualised models and the global model \cite{nguyen2022federated}.

 \end{itemize}

\subsection{The Healthcare Metaverse Platform}

\begin{figure*}[h!]
	\centering
        \includegraphics[width=0.9\linewidth]{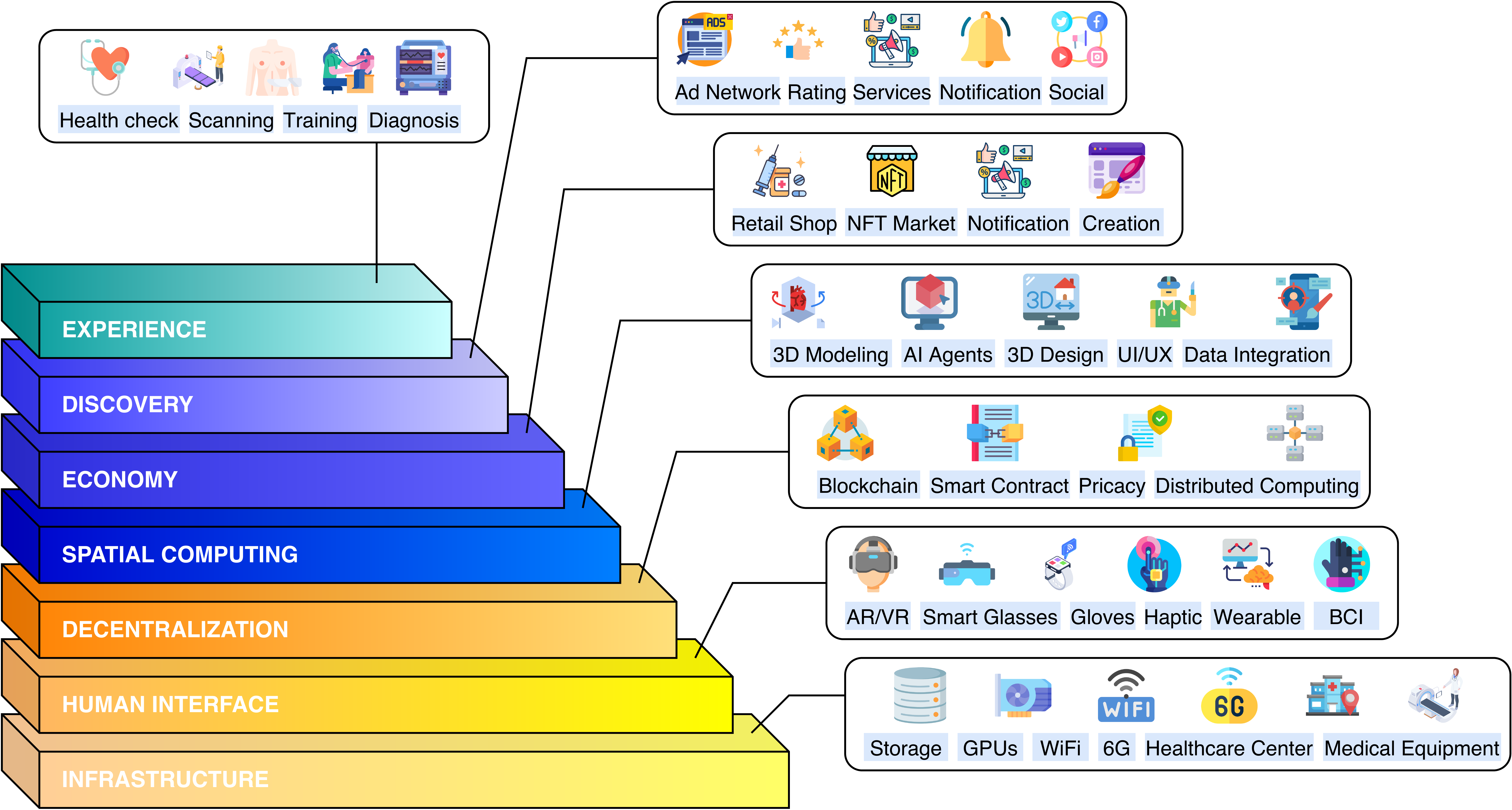}
	\caption{The Healthcare Metaverse Platform.}
	\label{fig:platform}
\end{figure*}

Metaverse has been built to intently provide permanent, decentralized, collaborative, and interoperable opportunities and business models that will allow companies to extend their digital operations \cite{han2023application}. A general Metaverse platform for healthcare and medical can have seven layers, including infrastructure, human interface, decentralization, spatial computing, economy, discovery, and experience as from the bottom to the top layers as shown in Fig. \ref{fig:platform}, as from the bottom to the top, but not being strict and may be changed upon many metrics, such as applications, services provided, scalability, and requirement.

\textit{Infrastructure}: At the bottom, the infrastructure layer pertains to all technological infrastructure things that are required to create a fully immersive, functional, and interoperable Metaverse \cite{rawat2023metaverse}. Networking and wireless communication with 5G, 6G, and WiFi, cloud storage architecture, VR systems, microelectromechanical systems, and graphics processing units are some major technology clusters to build a healthcare Metaverse, which allows to seamlessly connect decentralized healthcare and medical centers. So far the Metaverse stands on the foundation of powerful computers, integrated medical circuits, communication components, healthcare and medical equipments, and mixed reality devices.

\textit{Human interface}: The human interface layer mentions about the hardware or the devices that enable users (doctors, clinicians, and patients) to experience the true immersive experience in the virtual world \cite{aslam2023metaverse}. This layer involves wearable devices such as VR mobiles, VR headsets, gloves with haptic, smart glasses, gesture, voice, miniaturized biosensors, smart contact lenses and human brain-computer interface systems. With the haptic technology, users can operate electronics gadgets in mid-air (e.g., picking up a 3D lung object to zoom in and rotate for diagnosis) without touching screen or button-aided control.

\textit{Decentralization}: In the domain of healthcare, the Metaverse should be decentralized, open, and diffused, that means, it is governed by a single entity and belongs to no one but serving everyone at the same time. Decentralization encompasses the blockchain, self-sovereign digital identity mechanisms, smart contracts, open-source platforms to enable and facilitate the increasing growth of distributed computing and healthcare microservices in a scalable environment, serving many crowded patient and medical expert communities \cite{villarreal2023blockchain,huynh2023blockchain}. A host of decentralized healthcare Metaverse projects can use the blockchain technology to provide patient-owned private data (e.g., sensory data, medical image, and electronics health records) and shareable clinic experiences (e.g., diagnosis and treatment approaches) in a transparent and traceable manner to perform transactions and interactions.

\textit{Spatial computing}: as standing for a real/virtual combination, spatial computing dims the borderline between the actual and ideal worlds. In the healthcare domain, spatial computing is all material, libraries, and AI tools to create 3D immersive experience for real-world healthcare and wellness applications as well as medical services \cite{gupta2023understanding}. For instance, medical students have opportunity to practice and test their ability in realistic environments without risk over a virtual surgical training applications built in the Metaverse. Some 3D design engines like Unity and Unreal support for 3D modeling and rendering with GPUs for displaying geometry and animation. Spatial computing also helps to integrate diversified healthcare and medical data from different devices and sources for between visualization and analysis using VR devices.

\textit{Economy}: This layer is responsiable for economic establishment and development in the Metaverse, in which patients, doctors, healthcare institutions, and medical clusters can manage their business in many unique kinds. Patients can allow the service providers to use their private data for research activities and other non-commercial purposes, meanwhile, the providers will reward their contribution by virtual goods, assets, and incentives. There are some economic actitivies provided by healthcare and medical institutions, such as managing a virtual market of digital contents (e.g., medical training, remotely collaborative diagnosis, and treatment plan) and trading health things and medical equipment (e.g., first aid kit, thermometer, and wheelchair), wherein retail shops should be placed in the virtual world \cite{huang2023standard}. Indeed, it has enormous potential for economic expansion by generating a virtual space for not only patients and doctors but also content creators and designer to connect and communicate in the Metaverse.

\textit{Discovery}: Discovery can be understood as an advertising ecosystem which includes store placements, rating systems, and recommendation systems from users and service providers. This layer conveys the experiential learning that can be achieved as the results of pull and push information constantly. Particularly in the Metaverse for healthcare and medical domains, pull represents an inbound system where patients proactively look for useful information and experiences (e.g., disease symptoms and treatment plans shared by other patients), whereas push in more outbound and incorporates notice procedures about what experiences awaiting patients in the Metaverse \cite{wu2023into}. Some major modules are usually deployed in the Metaverse to make profit, including inbound with search engine, community-driven content, application stores, and real-time presence, and outbound with display advertising, notifications, emails and other social media.

\textit{Experience}: This layer contains what the most individuals and business are focusing on. Users, including patients and clinicians, has various interactive activities, with virtual objects and each other, in digitally-driven setting via digital content such as disease symptom declaration, health and medical checking, remotely collaborative diagnosis, medial training and education. As known that the Metaverse is more than a simple 3D representation of reality, it was introduced as the ultimate virtualization of physical space and product to engender a seamlessly virtual-realistic ecosystem with built-in services and applications. For example, an interactive virtual reality surgery platform can help doctors to find an anatomy of interest and utilize meaningful information, from learned AI model, to enhance the quality of medical images in the real time.

\section{Benefits of FL in Healthcare Metaverse}
\textcolor{black}{This section highlights the potential benefits of adopting FL in the healthcare Metaverse environment such as providing improved privacy, better interoperability, data management and cross-model usage, extra security, automation and intelligence, improved scalability and low latency support.} 


\subsection{Improved Privacy}
The Metaverse provides doctors and patients with an immersive and interactive healthcare experience with the aid of technologies such as AI, AR/VR/MR, blockchain and digital twins. The Metaverse environment can collect the patients’ details such as the brainwaves, health related data, and biometric data to create a complete healthcare system that simulates the healthcare system of the physical world and could use the same for understanding the patients’ body, their physiological responses, and the way their body would react in case of specific treatment procedures. However, such data, if fallen into the hands of the malicious users could manipulate it, thus leading to serious complications such as improper diagnosis, treatment, and drug administration. Therefore, preserving data privacy is of great importance in the case of healthcare Metaverse \cite{zhang2023multi}. FL is the privacy-preserving machine learning mechanism that enables different participating entities to jointly collaborate a machine learning model without actually sharing the data, but only the model parameters. As most of the data of a particular patient is maintained by a single edge device/ hospital, FL mechanisms help in training the data on the edge devices itself. This enables the centralized medical agencies to improve their models without actually collecting the private data of users. 
\subsection{Better Interoperability between Different Stakeholders}
The healthcare Metaverse can collect data from multiple healthcare centres/ hospitals for ensuring greater patient diversity from different demographics \cite{wischgoll2023metaverse}. However, data interoperability between different stakeholders is a significant challenge in the Metaverse environment, since the data collected from multiple devices/ Metaverses need to be made accessible to the trusted parties. Such data can be used for finding patient similarity \cite{pai2018patient}, remote patient monitoring \cite{coffey2021implementation}, and for predicting mortality rates. In order to address these issues, FL can be used for pooling data from every edge node/ hospital, train them on the edge devices itself and send the updates to the server, so that the model can be globally trained in the Metaverse, thereby mitigating localized biases and offering customized and effective solutions for rare diseases or during a pandemic. The problem of data interoperability can thus be addressed using FL mechanisms, thereby providing quality healthcare to the patients. 
\subsection{Better Data management and Cross-model usage}
The data sources of the healthcare Metaverse include sensors, AR/VR devices, wearable devices, scanning machines and other healthcare components. Such data that is widely distributed needs to be aggregated for making intelligent decisions in a healthcare scenario. This is particularly useful in conducting collaborative research in applications such as building a unified and global model for disease diagnosis, identifying disease pattern trends and so on. However, due to the heterogeneity of data, it becomes quite challenging to train the data at the central server \cite{zytko2023dating}. FL mechanisms can be employed here to deal with the issues of data heterogeneity by training the data at the devices itself. Vertical FL, also known as heterogeneous FL helps in training the datasets that share the same sample ID space but differing in feature space. 
Thus FL mechanisms can guarantee proper data management and cross-model usage witin and between different healthcare Metaverses.

\subsection{Extra Security}
The major security challenges in the Metaverse environment include user identity spoofing and account hacking, thus resulting in the avatar itself to be taken over by the intruders. It is critical in the case of a healthcare Metaverse, as the intruders can use such data collected from hospitals, patients and other healthcare providers to manipulate the data itself or the decisions regarding disease diagnosis and medications to be provided \cite{beck2023educational}. FL can help in such scenarios by making use of its secure aggregation mechanisms to keep the updates of the edge devices private. The actual value or the source of model updates is thus secured, by reducing the likelihood of data attribution and inference attacks. FL mechanisms can thus ensure extra security in a healthcare Metaverse. 

\subsection{Automation and Intelligence}
The healthcare Metaverse enables medical professionals to provide quality services to the patients, which was difficult earlier especially due to geographical limitations. With the advancements in the key enabling technologies such as AR/VR, AI, DT, and blockchain, automation in the healthcare Metaverse is possible in all the application areas such as diagnostics, billing, scheduling of appointments, data sharing, asset tracking and post-treatment care. However, the automation process needs to be energy-efficient and faster \cite{tlili2023metaverse}. As the FL model doesn't require collecting and aggregating data from different edge devices, the entire learning process can be accelerated, thus enabling enhanced energy-efficiency and intelligent responsiveness in terms of automation and relevant decision making. Thus, the comprehensive, intelligent and automated healthcare scenario in the Metaverse can be significantly improved with FL by learning from every client, yet by keeping the data private. 

\subsection{Improved Scalability}
The healthcare Metaverse signifies an immersive and shared environment that spans a myriad of 3D virtual worlds of patients and healthcare providers. However, building a scalable Metaverse is quite challenging as it is almost impossible to use traditional servers \cite{mcstay2023metaverse}. The scalability issues in a healthcare Metaverse correspond to the multitude of users interacting in the virtual and real worlds, complexity of the scene with respect to its appearance and detailing, and the range of interactions between the users. The healthcare Metaverse architecture should be built in such a way that the system efficiency is not compromised and best personalized experience could be provided to the patients and the healthcare providers. FL techniques can help in solving the scalability issues in the healthcare Metaverse by providing a de-centralized system for various medical applications such as diagnosis and treatment. This de-centralized system allows an efficient and concurrent system usage for a wide variety of users in the Metaverse healthcare scenario. 


\subsection{Low Latency Support}
\begin{figure*}[h!]
	\centering
	\includegraphics[width=\linewidth]{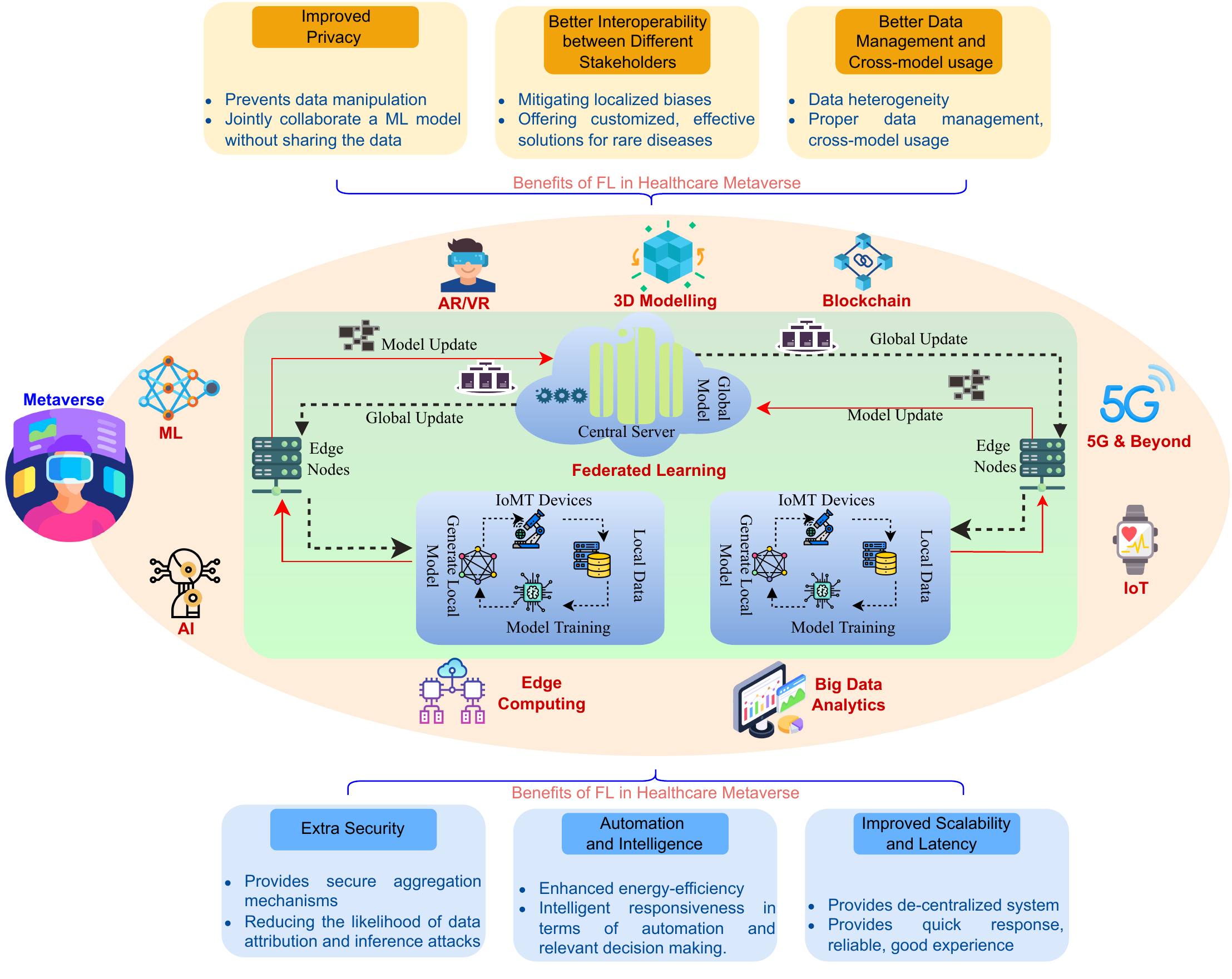}
	\caption{Benefits of FL in the healthcare Metaverse.}
	\label{fig:FL benefitss}
\end{figure*}

Network latency refers to the delay in communication over a network. In a healthcare Metaverse, network latency plays a major role especially in critical applications such as surgeries that require quick response to deliver a safe, reliable, and good experience. Any delay in communication could have serious effects leading to life-threatening situations. In order to support immersive Metaverse experiences, approximately 0-20 millisecond latency is required. Healthcare Metaverses require even lower latency for supporting services such as personalized medical recommendations, collaborative surgeries and quick-reaction services between patients an doctors present at various locations \cite{gruson2023new}. FL support collaborative model training without having the data to be shared to the centralized server, thus minimizing the amount of data transfers required between the edge devices and the healthcare Metaverse environments. FL thus help in personalized healthcare recommendations, disease diagnoses and other prediction and recommendation systems in a healthcare Metaverse environment. Fig. \ref{fig:FL benefitss} summarizes the benefits of FL in the healthcare Metaverse.
\section{Applications}
\textcolor{black}{The various applications pertaining to FL enabled healthcare Metaverse are discussed in this section. The healthcare applications explored are medical diagnosis, patient monitoring, collaborative research, medical education, infectious diseases or pandemics and drug discovery. }

\subsection{Medical Diagnosis}
Medical diagnosis deals with identifying patients' diseases based on their symptoms \cite{sun2021pmrss}. The conventional method of disease diagnosis is carried out physically at health centres or hospitals, where the patient needs to be examined by a doctor. However, during emergencies, the conventional diagnosis approaches may result in delayed treatment that may cause loss of lives. Therefore, telemedicine was introduced as a solution for monitoring patients remotely \cite{romanovs2021challenges}. Furthermore, the rapid developments in AI have contributed to all domains, including healthcare. By mimicking human cognitive capabilities, ML models can make better decisions regarding disease diagnosis by getting trained on the data collected from various health centres \cite{aggarwal2021diagnostic}. However, one of the notable limitations of telemedicine is that the patients cannot clearly express their symptoms or the level of injury as they don't meet the health professionals in person. 

The Metaverse is an ideal solution in such scenarios where enabling technologies like AR, VR, digital twins, and blockchain helps create a real-world digital simulation. Medical professionals can even provide personalized diagnoses with the help of the Metaverse, where doctors and patients can interact in shared, persistent and immersive virtual worlds \cite{qiao2022sprechd}. ML models can use the data collected using wearable devices and other systems in the Metaverse environment for training the model. However, this requires the entire data to be present in the system where the training is done. In addition, as the data collected may contain patients' private information, any leakage of such information may have serious consequences. For instance, the Metaverse environment requires data from the real world to create digital simulations of the patients through digital twins \cite{alazab2022digital}. Digital twins can be created for the patient population, individuals, or the organs. These digital twins can be manipulated to obtain clear insights in the decision making process. The data may be collected through different sources such as AR/VR devices and wearables. If the malicious users get hold of this collected data, they could manipulate it to create false diagnoses and improper treatments. Therefore, the model training should happen without exposing the patients' private information. 

\textcolor{black}{Consider an example where a unified global disease diagnosis model need to be built for the healthcare Metaverse. The centralized healthcare Metaverse requires the data to be collected from various edge devices or other Metaverses. However, as the patient's data is sensitive, sharing the data in its raw form to the central server for training the global model may violate the HIPAA rules \cite{annas2003hipaa}. However, a machine learning model cannot be trained without having the actual data made available to the original model. FL helps in such scenarios by creating a global disease diagnosis model without sharing the original data as such. In the case of horizontal FL, all the datasets across devices use the same set of features \cite{antunes2022federated}. Initially, a global model would be created and broadcasted to the edge devices or the participating Metaverses. The model would then be trained locally at their side with their own datasets. Only the parameter updates pertaining to the model will be sent back to the central server. After averaging the updates from different clients, the server updates it's original model and the updated model will be sent again to all the participating clients.This process continues until a predefined set of rounds or until optimal results are obtained. As multiple devices or Metaverses can jointly train a model under the coordination of the central Metaverse server while retaining the training data at the client side itself, FL mechanisms play a major role in ensuring data privacy in the Metaverse environment \cite{thomason2021metahealth}.}

\subsection{Patient Monitoring}
The Metaverse has completely revolutionized remote patient treatment and monitoring. The use of blockchain, AR, VR, MR and decentralized applications provide solutions to the problems of designing and developing Metaverse. This Metaverse would offer real-time patient monitoring ensuring rich human interaction and recreation of in-person experience. Immersive 3D interfaces are used in this regard that provide vivid and realistic user experience for monitoring of patients from geographically distributed locations enabling synchronized and real-time digital interaction between patients and healthcare providers. Telemedicine in healthcare is one aspect that helps in providing medicine as a remote service. The need of telemedicine has been felt significantly during the surge of COVID 19 \cite{portnoy2020telemedicine}. Before the pandemic of 2020,  43 percent of healthcare facilities provided remote treatment to patients and now the figure stands at 95 percent \cite{koonin2020trends}. The use of Metaverse technologies have curtailed the need of routine physical consultations by doctors and nurses which have further reduced their respective workload providing the ability to treat minor health conditions remotely. Telemedicine consultations through VR have open possibilities to interact with consultants and experts from any distant places across the globe by the use of simple head sets. The MRI and scans can be performed locally in a nearby facility and the data is transferred to the medical expert located anywhere in the world. This type of facility is extremely beneficial in countries that face acute shortage of medical professionals. On the contrary, patients from rural and remote regions need not travel greater distances to avail medical facility \cite{chengoden2022Metaverse,bansal2022healthcare}.

It is important to mention the role of Digital Twin in this regard, wherein test dummies could be used to understand how a patient would be recovering from a surgery or how a post operative patient would react to a specific medication \cite{erol2020digital}. Metaverse also empowers the use of medical wearables for patients and healthcare providers wherein these devices provide alerts to emergency staff and care givers to alert them in case of exceptional events like seizers, Chronic Obstructive Pulmonary Disease flare-ups or any healthcare anomalies for remote patients. The wearables enable physicians to get high quality data and make necessary inferences and predictions for accelerated and immediate dissemination of treatment through AI. These technologies also facilitate Metaverse aided daily check-ups enabling necessary decision making.  In case pf Psychiatric patients Metaverse based remote healthcare treatment is highly effective which help them interact in situations which cause anxiety and depression. The use of AR, VR based Metaverse helps healthcare professionals to closely monitor the remote environment of such patients and alert family members on situations which might cause anxiety, anger or psychosis attacks \cite{song2022Metaverse}.

\textcolor{black}{The use of FL is significant in this regard wherein secured data transfer across the stake holders is necessary to ensure seamless rendering of remote treatment facility. The traditional systems which store and maintain healthcare records in centralized computers are vulnerable to theft and data tampering. Since patients are remotely located, the use of FL provides distributed platform of intelligent techniques enabling multi-institutional collaborations ensuring privacy of patient data. In case of remote monitoring, the use of FL provides increased network capacity which also support Quality of Service (QoS). This enables patients to get enhanced experience wherein medical professionals can seamlessly monitor patients and get information on any anomalies immediately without delay ensuring immediate and accelerated treatment. Also, with the ever increasing population and scarcity of healthcare experts the use of FL provides the scalability to handle the huge patient load wherein considerable number of patients are treated remotely through the Metaverse relieving burden of the healthcare professionals \cite{rieke2020future}. }

\subsection{Collaborative Research and Medical Education}
Digital transformation has been adopted enthusiastically in various domains especially in collaborative research due to the emergence of virtual reality and related technologies. Further, VR has visualized rapid proliferation with the progressive growth in Metaverse technology. The increased adoption of these technologies have reduced the associated cost of implementation and made it more accessible for common users \cite{ahuja2023digital}. Experimentation and research are intrinsic aspects of higher education programs  wherein researchers often find it difficult to learn and retain information without hands-on experimentation. Also justification of the experimental results lag authenticity due to the same reason. The VR based simulated labs provide the necessary experimental and hands-on learning experience. In case of biomedical research, students can perform various critical experimentation using chemicals in a VR based safe zone with no risks of harming property or human life \cite{riva2002virtual}. The challenges pertaining to availability of chemical or related resources also gets eliminated wherein experiments can be repeated multiple times to ensure justification of results. Also, the removal of risks ensues innovation and creativity among student minds to design and develop experiments which could lead to path-breaking results and discoveries contributing immensely towards the growth of science \cite{iwanaga2023really}. The impact of Metaverse can be a game changing experience in collaborative research wherein improved data driven decision making would be possible across chemical value chain existing in various laboratories across the globe. A holistic approach to digital twin can enable improvement of process and control-system engineering from the very early stages of any collaborative project design. The use of advanced simulation and modeling technologies can enable multiple teams located globally to work on common projects, develop steady state design models and then integrate them into dynamic models. These models can be tested through real-time dynamic simulations for supporting process engineering execution \cite{aburayya2023sem}. \textcolor{black}{Although Metaverse enables such collaborative research but there are associated concerns pertaining to distributional discrepancies across multiple stake holders which could lead to counterproductive consequences. FL enables achieving collaborative equilibrium by ensuring security is not compromised wherein smaller collaboration coalitions can be created. In such cases each client can collaborate with other team members who maximally improve the model and isolate the ones whose contribution is minimum. FL enables multi-institutional collaborations ensuring data privacy is maintained. The model-learning manages all the available data without sharing it across the participating Institutions, The model-training is distributed across all the data-owners and the results are aggregated \cite{suzuki2020virtual}.} 

In the field of medical education, the Metaverse allows healthcare professionals to engage with each other through simulation based training while pursuing their medical education. As an example, in case of surgery, Metaverse usage need to be complemented by understanding of instrument usage that enables dexterous grasping capabilities. Thus the need evolves for the tracking of technologies being used to enhance the flexibility and adaptability of the Metaverse technologies. The doctors and experts can use the Metaverse technology to train fellow doctors and other medical staffs wherein virtual reality can enable trainers to get detailed visualization of the human body and also provide 360 degree view of the patient ailments. Similarly medical procedures could also be replicated using IoT, Digital Twins, 6G networks, robots, edge, cloud, quantum computing for the development of VR, AR and MR based systems to be used in medical diagnosis and treatment \cite{mozumder2022overview,siniarski2022need}. In case of surgeries, the use of robotics in the Metaverse environment would enable surgeries to be performed with enhanced precision and flexibility. The cloud based and real-time communication technologies in the Metaverse enables medical students to practise their skills with optimum level of accuracy ensuring patients get satisfactory healthcare service at the right place and time. Different types of assets in the form of avatars, 3D models, mixed reality and spatial settings are used in the Metaverse environment as an integrated package. As the number of students grow, the data and communication standards are also expected to grow rapidly facilitating cross-Metaverse communication. \textcolor{black}{The use of FL helps in simpler transfer of data across various platforms and networks ensuring privacy, scalability and interoperability. But dissemination of such services requires sharing of extensive medical image data across multiple Institutions which are used by healthcare professionals and researchers to develop data-driven models for delivering enhanced healthcare services. The amount of data available at individual healthcare centers are limited and deep learning models which are trained on the local centers alone fail to deliver the required level of performance. The best solution is to collect data from these individual centers and collate it in one center but this has associated challenged of compromising data privacy, The use of privacy-preserving algorithms on the data available at multiple centers would also serve the similar purpose of preserving privacy. The use of FL in such circumstances would enable deployment of large scale ML models trained on healthcare data available at individual centers without sharing of sensitive information. The FL framework would eliminate the need of transferring data and instead a general model could be trained on local data set which could be further transferred between the data centers \cite{sandrone2022medical}.}

In medical education, quality of training is dependent on variability of patients and the number of difficult cases handled in a particular health center. Considering geographical locations, the patient trait differ significantly across hospitals  wherein tertiary hospitals witness more volume in the number of difficult cases in comparison to secondary hospitals. The evaluation on an atypical patient at a certain hospital on the basis of such training model may be inadvisable as the model lags the experience to learn from previously handled atypical cases. A FL based model in the healthcare Metaverse would enable incorporation of data from multiple Institutions in association with virtual presence of healthcare experts leading to generation of accurate treatment to atypical patients even in secondary healthcare centers. 
\textcolor{black}{The exams conducted for medical students can also be verified from recordings of the procedures performed. This would enhance quality and standard of evaluation wherein a medical Institute expert faculty located in a geographically distant location can evaluate performance of the students using virtual reality or related technology in the Metaverse. Similarly, interactive question answer sessions can be conducted by using a VR to take an examiner inside the human body and ask questions on details of human anatomy or disease diagnosis. However, interoperability, scalability and privacy may be a concern. With the increase in the number of medical Institutes and number of students, such interactive exam sessions can confront potential latency issues. Also, since it is an examination, data privacy also acts as a challenge. The role of FL would enable negating all such challenges and help healthcare professionals to conduct exams without sharing of data. Also, the decentralized system in FL would enable multiple medical students to pursue such aforementioned interactive examinations effectively and concurrently in a Metaverse healthcare environment \cite{huh2022application}. }

\subsection{Infectious diseases/ Pandemic}
The COVID-19 pandemic has set up a perfect stage for the Metaverse technology to take off. The pandemic has affected the economic and social stability of the nations to a great extent. Longer periods of isolation has even created a society with social anxiety \cite{buck2020ecological}. However, the pandemic has made huge technological advancements in education, healthcare, and entertainment. The wide adoption of online classes, virtual meetings, telemedicine and OTT movie release are all a result of the pandemic \cite{ifdil2023virtual}. Pandemic surveillance is an important aspect that helps to understand the pattern of the spread of the disease so that appropriate measures can be adopted to alleviate the spread \cite{onggirawan2023systematic}. The transparency and efficiency of the surveillance can be enhanced by building a platform that can integrate data from multiple domains such as transportation, retail and telecommunications. However, such an integration would be difficult to be accomplished in a real world even with the advanced technologies due to the risk of the spread of disease. 

The Metaverse can be adopted here by creating a digital world replica of the real world and understand how the real world acts and reacts. Such a Metaverse realization can better detect a pandemic and respond effectively. However, as the data collected from different edge devices are heterogeneous, efficient cross-model usage becomes a significant concern. Also, security and privacy plays a major role in such scenarios, as the data needs to be shared among nations for creating a global disease prediction and pattern evaluation model to take any significant steps for mitigating the spread of the disease. Even healthcare Metaverse proves to be beneficial in performing surgeries \cite{van2022robotic}, patient monitoring and disease diagnosis \cite{yang2022smart}. FL can be adopted in such situations, as it provides extra security, better data management and cross-model usage. The authors in \cite{dou2021federated} used FL to detect COVID-19 related Computed tomography (CT) abnormalities. \textcolor{black}{As an example, considering the COVID-19 pandemic period, various countries had patients
suffering from different variants of the corona virus. Hence,
the relevant drug solutions would also differ based on the
location and type of the virus. Also, the kind of preventive measures to be adopted by each nation differ significantly. Hence, personalized solutions are required especially during pandemic period. FL enabled healthcare Metaverse helps in such situations by creating a global model that could be used by any of the participating clients for training and testing their own data. This would help to obtain customized solutions for the end user. Adopting FL in such situations will not only help in data privacy preservation, but also in addressing the scalability concerns in a Metaverse healthcare environment.}The technique proved to be efficient in preserving patients' privacy, when compared to the traditional method of training the data locally at the edge devices. FL thus helps in building a global model without having the original data to be shared to any centralized Metaverse server, thereby securing the data from any kind of malicious users.
\subsection{Drug Discovery}
Drug discovery is the technique through which potential novel therapeutic entities are identified, using an amalgamation of experimental, computational, clinical and translational models. Machine learning models play a major role here in improving the decision making for applications such as hit discoveries, target identification, hit to lead and lead optimization \cite{dara2022machine}. Computational growth algorithms are also used in De novo drug design (DNDD) to understand and design new chemical entities that fits in the defined constraint sets \cite{mouchlis2021advances}. Enhanced experimental results can be obtained if researchers can collaborate in building the required models. However, conducting experiments for discovering new drugs is critical and requires a lot of human effort and expenses. 

Healthcare Metaverses can help in such scenarios by creating a virtual simulation of the real-world laboratory or by creating  a completely new experimental laboratory space where multiple hospitals and researchers can experiment on drug design either by sharing the original data set to the central medical agency or by sharing only the intermediate results \cite{joshi2023collaborative}. However, such experimental research also face significant challenges as explained below. Preserving data privacy is challenging if the data is shared in its raw form. If the participating entities decide to share only the intermediate results, the drug discovery process may not succeed always as most of the process would be hidden from the central laboratory. Also, scalable solutions are required in such scenarios, as the existing approaches may not work well to accommodate such large data. 
\begin{figure*}[h!]
	\centering
        \includegraphics[width=\linewidth]{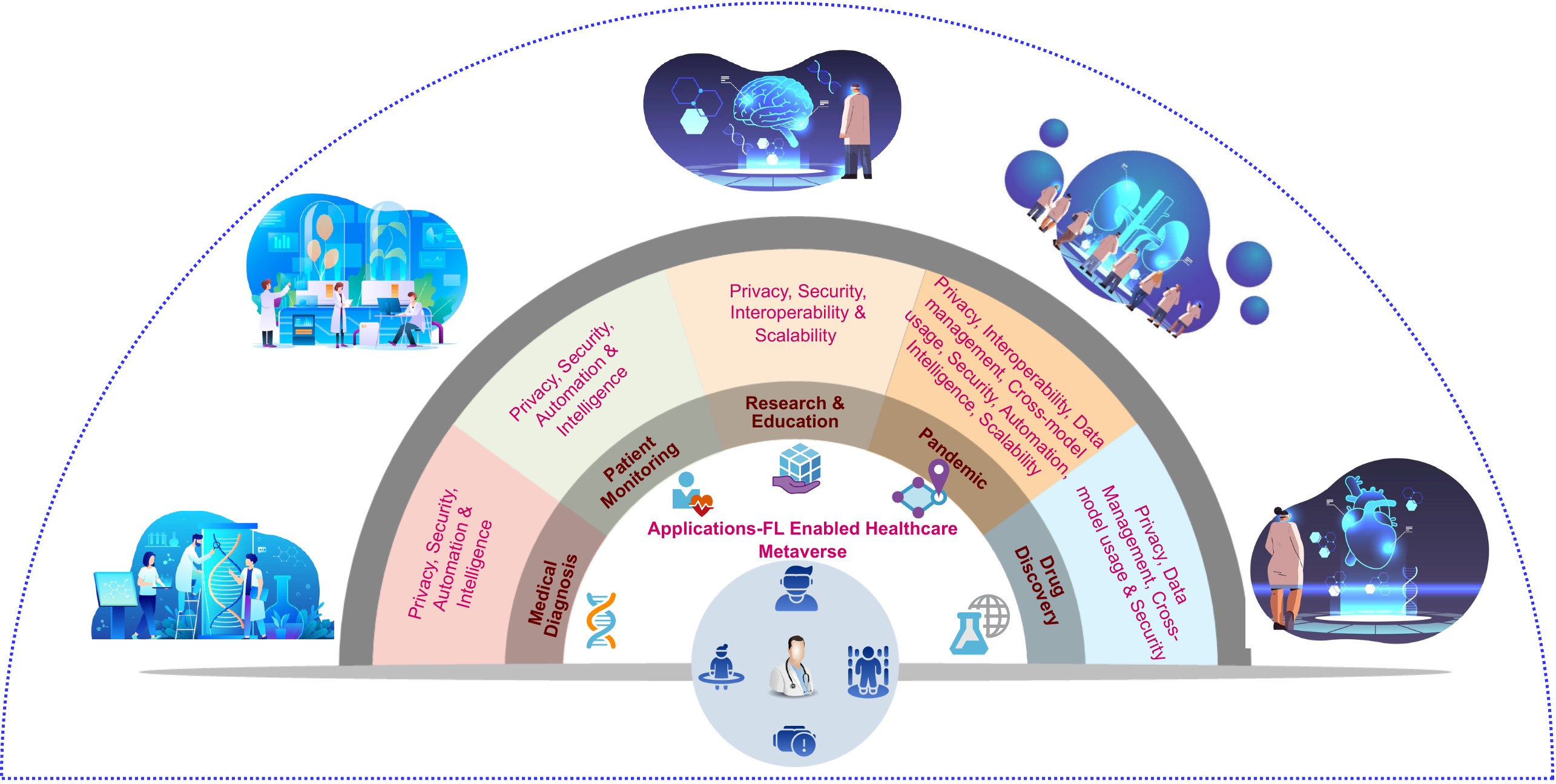}
	\caption{Applications of the FL-enabled Healthcare Metaverse.}
	\label{fig:FL applications}
\end{figure*}
FL can be applied in the healthcare Metaverse for ensuring data privacy and scalability in the drug discovery process. This helps in training a global model for decision-making across various critical applications such as  prediction of the 3D- protein structure, drug-protein interactions, analyzing the efficiency of a drug, and guaranteeing safety biomarkers within and between Metaverses. \textcolor{black}{Consider a scenario where various countries have agreed to develop a global drug prediction model with the data from the geographically distributed patients \cite{shi2023new}. The relevant drug solutions would differ based on the location and type of the disease. In such circumstances, research related to drug discovery of the disease would require data pertaining to all of such variants rather than being skewed to a particular location or community. Hence cross-institutional data is required to enable exhaustive research on the virus leading to the discovery of effective drug solutions targeting each variant of the deadly virus. The role of FL is significant in such scenarios which enable secure transfer of such confidential data pertaining to molecular dynamics of the virus across multiple Institutes located at geographically distributed locations \cite{surveswaran2023glimpse}. The scalability and interoperability of the FL based Metaverse framework would enable scientists across the globe to collaboratively work on development of appropriate drug solutions based on cross-institutional and cross-national big dataset yielding customized and efficient solutions.} Another important application is in understanding molecular features that help in predicting the cancer drug response. The authors in \cite{chen2021fl} developed a horizontal FL model for Quantitative Structure-Activity Relationship (QSAR) analysis. As each of these applications are critical and require scalable solutions, FL proves to be the effective solution for dealing with such issues in the healthcare Metaverses for the drug discovery process. Fig. \ref{fig:FL applications} depicts the applications of the FL-enabled healthcare Metaverse. Table \ref{tab:technical} highlights the technical aspects of FL for various healthcare Metaverse applications.

\begin{table*}[h!]
\centering
\caption{Technical Aspects of FL for Healthcare Metaverse Applications.}
\label{tab:technical}
\begin{tabular}{|l|l|l|p{2.4cm}|l|p{1.8cm}|l|}
\hline
Application &
  Privacy &
  Interoperability &
  Data management   \& Cross-model usage &
  Security &
  Automation \& Intelligence &
  Scalability \\ \hline\hline
Medical Diagnosis &
  \cellcolor[HTML]{d1f8d1}H &
  \cellcolor[HTML]{FFFFC7}L &
  \cellcolor[HTML]{FFCCC9}M &
  \cellcolor[HTML]{d1f8d1}H &
  \cellcolor[HTML]{d1f8d1}H &
  \cellcolor[HTML]{FFFFC7}L \\ \hline
Patient Monitoring &
  \cellcolor[HTML]{d1f8d1}H &
  \cellcolor[HTML]{FFFFC7}L &
  \cellcolor[HTML]{FFCCC9}M &
  \cellcolor[HTML]{d1f8d1}H &
  \cellcolor[HTML]{d1f8d1}H &
  \cellcolor[HTML]{FFFFC7}L \\ \hline
Collaborative Research and Medical Education &
  \cellcolor[HTML]{d1f8d1}H &
  \cellcolor[HTML]{d1f8d1}H &
  \cellcolor[HTML]{d1f8d1}H &
  \cellcolor[HTML]{d1f8d1}H &
  \cellcolor[HTML]{FFCCC9}M &
  \cellcolor[HTML]{d1f8d1}H \\ \hline
Infectious diseases/ Pandemic &
  \cellcolor[HTML]{d1f8d1}H &
  \cellcolor[HTML]{d1f8d1}H &
  \cellcolor[HTML]{d1f8d1}H &
  \cellcolor[HTML]{d1f8d1}H &
  \cellcolor[HTML]{d1f8d1}H &
  \cellcolor[HTML]{d1f8d1}H \\ \hline
Drug discovery &
  \cellcolor[HTML]{d1f8d1}H &
  \cellcolor[HTML]{FFCCC9}M &
  \cellcolor[HTML]{d1f8d1}H &
  \cellcolor[HTML]{d1f8d1}H &
  \cellcolor[HTML]{FFCCC9}M &
  \cellcolor[HTML]{FFCCC9}M \\ \hline
\end{tabular}%
\begin{flushleft}
\begin{center}
    
\begin{tikzpicture}

\node (rect) at (0,2) [draw,thick,minimum width=1cm,minimum height=0.7cm, fill= yellow!35, label=0:Low Significance] {L};
\node (rect) at (5,2) [draw,thick,minimum width=1cm,minimum height=0.7cm, fill= red!25, label=0:Medium Significance] {M};
\node (rect) at (10,2) [draw,thick,minimum width=1cm,minimum height=0.7cm, fill= green!20, label=0:High Significance] {H};

\end{tikzpicture}
\end{center}

\end{flushleft}

\end{table*}

\section{Challenges and Future Directions}

In this section, we will discuss several challenges and future directions of FL in the healthcare Metaverse.

\subsection{Limited Computing Capabilities}
\subsubsection{Introduction to issues}
FL enables the healthcare Metaverse to develop a global model by gathering information from several decentralised healthcare clients in the Metaverse. FL allows on-device training in the healthcare Metaverse  and updates the global model based on local model updates while preserving the client's private local data. FL provides a lot of advantages, including scalability and data privacy in the healthcare Metaverse. However, healthcare IoT-enabled devices that are used in the healthcare Metaverse, such as surgery robots and low-cost computing healthcare devices built using limited processing power, bandwidth, and storage. This creates a challenge for adapting FL in the health Metaverse. 
\subsubsection{Possible Solutions}
Edge computing is a potential solution in overcoming this current challenge. Applications needing high bandwidth and low latency near the data source benefit from mobile edge computing. For instance, to save energy and compute resources, devices might offload difficult activities to nearby edge servers. Edge servers will provide services with minimal latency since they are significantly closer to consumers than cloud servers, enabling the Metaverse to deliver a service. The use of hierarchical cloud native technologies offers high computational and storage resources. To enhance devices' performance, the appropriate combination of edge and cloud-based applications is required \cite{wang2022ai}.

\subsection{Synchronization between physical and virtual worlds}
\subsubsection{Introduction to issues}
A crucial problem in the healthcare Metaverse is the synchronisation of the real and virtual worlds. The creation of healthcare-based global-model for patient diagnosis or at the time of drug creation in the FL enabled healthcare Metaverse demands real-time data from diverse devices and sensors.  In order to deliver data  from the physical world to the FL enabled healthcare Metaverse, a network that is dependable with minimal latency is required.
\subsubsection{Possible Solutions}
The 6G network is a potential solution in addressing synchronisation of the real and virtual worlds. 6G operate in the terahertz band and have a peak rate of 1T b/s and a network with ultra-reliable and low-latency communication of less than 1 ms. The total quality of experience for FL enabled healthcare Metaverse is greatly improved by 6G wireless networks \cite{siniarski2022need}.
\subsection{Privacy and Security }
\subsubsection{Introduction to issues}
FL is able to safeguard each participant's privacy in the healthcare Metaverse since the model training will be done locally, with just the model parameters being sent to the FL server. However, sharing model updates with other users in the healthcare Metaverse during training might still provide adversaries or intruders access to private data. FL in the healthcare Metaverse  is also susceptible to problems with communication security including jamming and distributed denial-of-service (DoS) attacks. With regard to jamming attacks in particular, an attacker may send out strong radio frequency jamming signals to obstruct or interfere with the connections between mobile healthcare Metaverse devices and the main server. Such an assault may result in model upload/download errors, which would reduce the FL systems' accuracy and performance in the healthcare Metaverse .

\subsubsection{Possible Solutions}
The challenge of user data privacy in the healthcare Metaverse can be overcome with the use of differential privacy and collaborative training. The performance, or model accuracy, is reduced when these methodologies are used, however. They also demand a substantial amount of processing power from participating healthcare Metaverse devices. As a result, while developing the FL system, the trade-off between privacy assurance and system efficiency must be carefully considered. The problem of security can be overcome by using  anti-jamming techniques like frequency hopping, which includes transmitting one extra copy of the model update across several frequencies.

\subsection{Lack of Explainabilty}
\subsubsection{Introduction to issues}
The black-box nature of FL models hinders users from understanding the model's output. The unsupported predictions or conclusions of the black-box model might result in a catastrophic failure in the healthcare Metaverse in a complex situation. As in the case of a medical equipment that uses a visual recognition model to identify a patient's sickness. if the model is difficult to understand and the doctor is unable to assess the results. Clinicians in the healthcare Metaverse may make bad decisions because of the black-box model's false interpretation. An interpretable FL model must be developed to provide a greater range of healthcare Metaverse judgements.
\subsubsection{Possible Solutions}
The usage of Explainable AI (XAI) is a potential solution to the issue of creating explainable and interpretable FL models in the healthcare Metaverse. XAI ha a collection of techniques and strategies that enable stack-holders in the healthcare Metaverse to grasp and rely on the output and results generated by prediction models.

\subsection{Fair FL}
\subsubsection{Introduction to issues}
In Fl, tens of thousands of devices will be used to train a single global model that will be used across the healthcare Metaverse. In such a system, naive global model optimization may be unfair to certain devices if it generates disproportionate advantages or disadvantages for them. FL towards fairness is a need. FL requires a rational incentive system and equal resource allocation. A significant difficulty in the healthcare Metaverse is the equitable distribution of computational and communication resources.
\subsubsection{Possible Solutions}
The development of high quality-based incentive mechanisms encourages more devices to take part in the creation of global model in the healthcare Metaverse. As a result, these devices are able to earn greater rewards for the high-quality data they contribute and the resources they share, which ultimately leads to an improvement in the reliability of the model as well as the improved learning performance of FL models. Additionally, this encourages the development of fair FL models.

\section{Conclusion}
Given the lack of a survey on FL-enabled Metaverse healthcare, we have provided a detailed survey on the usage of FL for Metaverse healthcare. We have first presented the fundamentals of FL, healthcare systems, and Metaverse, as well as the introduction to the benefits of FL and Metaverse in intelligent healthcare. The key applications of FL in Metaverse healthcare in the open literature have been explored. One of the applications is medical diagnoses, where FL helps to create a collaborative and global model without sharing the original data from IoT devices. The key applications also include patient monitoring, collaborative research and medical education, infection prevention, and drug discovery, where the focus is to use the capabilities of FL in using data availability, exploiting powerful resources at distributed IoT devices, and improving data privacy. Based on the reviewed literature, we have emphasized vital challenges in using FL for Metaverse healthcare and pointed out several interesting research solutions. It is noted that both FL and Metaverse are new technologies, and there are lots of room for improvements in intelligent healthcare and smart medical services. We do expect that this work will stimulate more research attention and drive innovative research ideas for FL-enabled Metaverse healthcare.

\bibliographystyle{IEEEtran}
\bibliography{Ref}
\end{document}